\documentclass[oribibl]{llncs}
\usepackage{graphicx}
\usepackage{capt-of}
\usepackage{comment}
\usepackage[]{algorithm2e}
\usepackage{multirow}
\usepackage{url}
\usepackage[section]{placeins}
\usepackage{rotating}
\usepackage[tight,footnotesize]{subfigure}
\title{Multi-layer Architecture For Storing Visual Data Based on WCF and Microsoft SQL Server Database }

\author{Rafa{\l} Grycuk\inst{1}, Marcin Gabryel\inst{1}, Rafa{\l} Scherer\inst{1}, Sviatoslav Voloshynovskiy\inst{2}}
\authorrunning{Rafa{\l} Grycuk et al.}
\institute{Institute of Computational Intelligence, Cz\c{e}stochowa University of Technology\\
Al. Armii Krajowej 36, 42-200 Cz\c{e}stochowa, Poland\\
\email{ \{rafal.grycuk, marcin.gabryel, rafal.scherer\}@iisi.pcz.pl\\
\texttt{http://iisi.pcz.pl}}
\and
University of Geneva, Computer Science Department, \\
7 Route de Drize, Geneva, Switzerland \\
group web page: \texttt{http://sip.unige.ch}
}
\begin{document}

\maketitle 

\begin{abstract}
In this paper we present a novel architecture for storing visual data. Effective storing, browsing and searching collections of images is one of the most important challenges of computer science. The design of architecture for storing such data requires a set of tools and frameworks such as SQL database management systems and service-oriented frameworks. The proposed solution is based on a multi-layer architecture, which allows to replace any component without recompilation of other components. The approach contains five components, i.e. Model, Base Engine, Concrete Engine, CBIR service and Presentation. They were based on two well-known design patterns: Dependency Injection and Inverse of Control. For experimental purposes we implemented the SURF local interest point detector as a feature extractor and $K$-means clustering as indexer. The presented architecture is intended for content-based retrieval systems simulation purposes as well as for real-world CBIR tasks.

\end{abstract}
\keywords{WCF, Microsoft SQL Server, Dependency Injection, Inversion of Control, Entity Framework, Multi-layer Architecture, $k$-means, SURF, Content-Based Image Retrieval}
\section{Introduction}
Images are created everyday in tremendous amount and there is ongoing research to make it possible to efficiently search these vast collections by their content. Generally, this work can be divided into image classification \cite{JAISCR-optimized}\cite{DrozdaSG14} and image retrieval \cite{DrozdaSG13}. 
Recognizing images and objects on images relies on suitable feature extraction which can be basically divided into several groups, i.e. based on color representation \cite{Huang1997correlograms}, 
textures \cite{smietanski2010texture}, shape \cite{Veltkamp:2000Shape}\cite{zalasinski2013}, edge detectors \cite{ogiela2002syntactic}\cite{ogiela2005nonlinear}\cite{Zitnick:2014} or local invariant features, e.g. SURF \cite{SURF:Bay:2008}, SIFT \cite{SIFT:Lowe:2004} or ORB \cite{Rublee2011}. Matching features can be also performed by several methods, e.g. clustering, nearest neighbour, bag of features \cite{Grauman2005} or soft computing \cite{Cpalka2009e1659}\cite{Cpalka2013}. 

There are many content-based image processing systems developed so far. A good review of such systems is provided in \cite{veltkamp2002content}. To the best of our knowledge no other system uses similar set of tools to the system proposed in the paper. Now we describe briefly the most important tools used to design the proposed framework. 
\subsection{Windows Communication Foundation}
Windows Communication Foundation (WCF) is a framework based on Service-Oriented Architecture \cite{cibraro2010professional}\cite{hirschheim2010service}. WCF allows to send data asynchronously between two service and client endpoints. Service endpoints can be deployed on IIS server or be hosted locally. The messages can be send as XML (value types) or binary file (complex types) \cite{chappell2010introducing}\cite{klein2007professional}\cite{mackey2010windows}\cite{martin2006agile}. WCF consist of following features:\cite{rosen2012applied}\cite{tsai2010service} Service Orientation, 	Interoperability, Multiple Message Patterns, Service Metadata, Data Contracts, 	 Security, Multiple Transports and Encodings, Reliable and Queued Messages, Durable Messages, Transactions, AJAX and REST Support, Extensibility.

\subsection{SQL Sever}
MS SQL Server is a database  management system for storing various types of data, fully supporting cloud computing technologies \cite{ben2012microsoft}\cite{ChromiakS12}\cite{ChromiakS14}\cite{delaney2013microsoft}. It provides a set of tools to extract data from various devices or sources, even at datacenters. MS SQL query language, T-SQL (Transact-SQL), allows for both structural or procedural queries \cite{jaiscr4}\cite{jaiscr1}\cite{jaiscr2}\cite{leblanc2013microsoft}\cite{jaiscr3}\cite{jaiscr5}. The DBMS (Database Management System) is based on a client-server architecture. The platform is composed of the following services:\cite{lobel2012programming}\cite{mistry2012introducing}
\begin{itemize}
\item Database engine - allows to execute queries and is necessary to run the server,
\item Integration Services (SSIS, SQL Server Integration Services) -  ETL (Extraction, Transformation and Loading) platform responsible for data migration from the heterogeneous data sources,
\item SQL Agent -  answerable for performing tasks according to the specified schedule,
\item Full-text Filter Daemon Launcher - allow to perform full-text searches on text columns,
\item Reporting Services (SSIS, SQL Server Reporting Services) - responsible for designing and deploying reports,
\item Analysis Services (SSAS, SQL Server Analysis Services) - allows to create multidimensional cubes and executing MDX (Multidimensional Expressions) queries \cite{harinath2012professional},
\end{itemize}
DBMS store data in a relational form (tables and their relations) and allows to select information by executing queries. 
	Many frameworks perform object relational mapping. In this paper we use Entity Framework (EF) with Code First approach \cite{adya2007anatomy}\cite{castro2007ado}\cite{lerman2011programming}.
\subsection {Dependency Injection and Inverse of Control} \label{DI_IOCsub}
\begin{figure}[!htb]
 \centering
 \includegraphics[width=10cm]{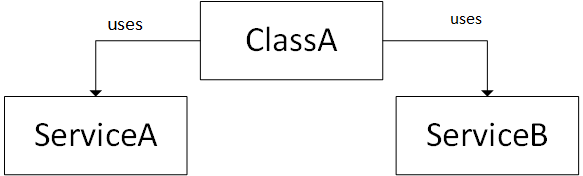}
 \caption{Problem illustration of tightly coupled dependencies. $ClassA$ uses $ServiceA$ and $Service B$. This is a simple representation of $ClassA$ dependencies  on $ServiceA$ and $ServiceB$.  } \label{DI_IO}
 \end{figure}

\begin{figure}[!htb]
  \centering
  \includegraphics[width=10cm]{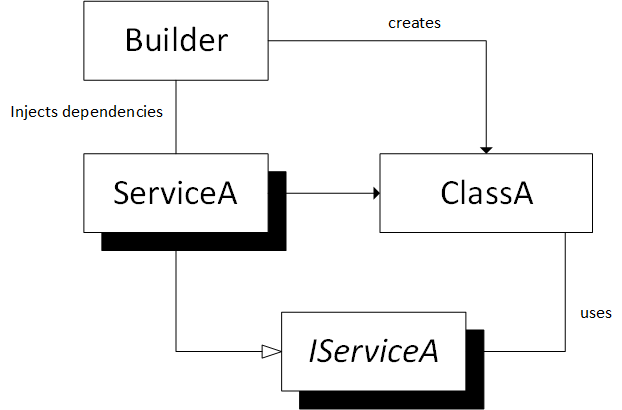}
  \caption{Solution of the problem formulated in Fig. \ref{DI_IO}. The Inverse of Control pattern can be implemented by Dependency Injection. $Builder$ creates $ClassA$ which uses abstract interface $IServiceA$. Dependencies of $ServiceA$ are injected to $ClassA$ by inheritance.  } \label{DI_IO2} 
  \end{figure}

Re-usability of existing components is crucial in modern software engineering. The aim of this approach is to combine separate layers into one application. This is a challenging task, because as the application complexity increases, so do dependencies \cite{caprio2005design}. The best practice for tone down proliferation of dependencies is by using Dependency Injection (DI) design pattern allowing to inject objects into a class constructor. Thus, the creation of the object does not rely on a class. The initialization logic is rarely reusable outside of created component. That pattern provides a layer of abstraction for the injected object, thus we can implement the concrete logic in the other component and inject it in the class constructor by the interface.  DI \cite{martin2006agile} is an implementation of Inverse of Control (IoC)\cite{fowler2004inversion}\cite{prasanna2009dependency}\cite{seemann2012dependency}. Figure \ref{DI_IO} shows the typical class dependencies. Such a scheme entails the following problems \cite{magazine2005design}:
 \begin{enumerate} 
 	\item Any code changes of $ServiceA$ forces changes in $ClassA$, thus the recompilation of all components is required,
 	\item All classes must be implemented and available at the compile time,
 	\item Classes are difficult to test and to achieve components isolation,
 	\item Contradicts the (DRY) Don't Repeat Yourself principle.
 \end{enumerate}
 To resolve this issue we used the Dependency Injection \cite{caprio2005design}\cite{fowler2004inversion}. A Conceptual view of DI is presented in Figure \ref{DI_IO2}. 
\section{Proposed Architecture For Storing Visual Data }
\begin{figure}
\subfigure[Multi-tier architecture for storing visual data. Final components are compiled release of dll's and open components are user implementation.]{\includegraphics[width=4cm, height=6cm]{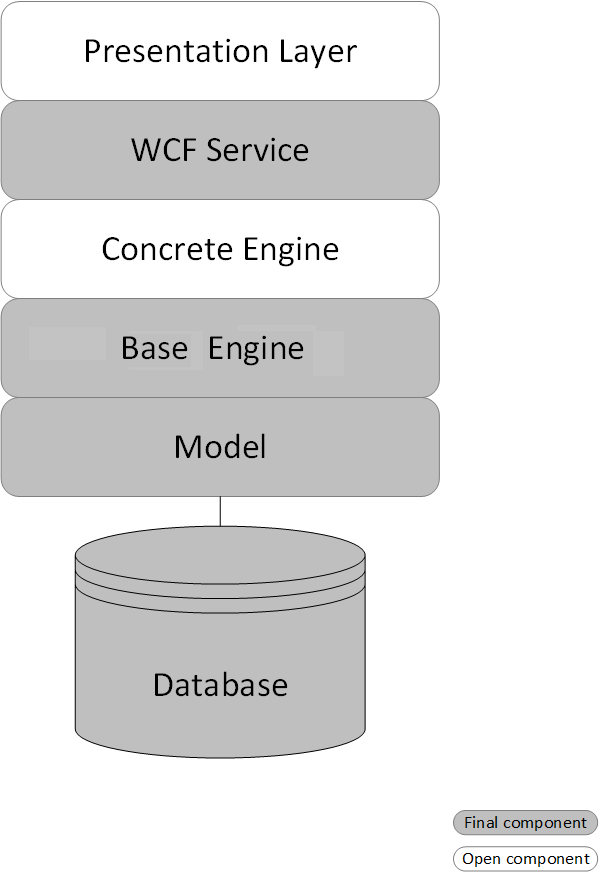}}
 \subfigure[Entity diagram based on database tables.]{\includegraphics[width=7.5cm, height=6cm]{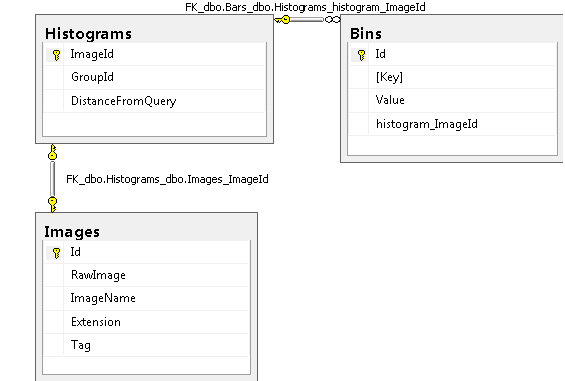}}
\caption{Multi-layer architecture and entity diagram of the proposed approach.  }
\label{layer}
\end{figure}

Proposed architecture  is based on two main components. The first one is SQL database which is Microsoft SQL Server, and the second one is Windows Communication Foundation (WCF). Our approach consists of five main layers (tiers):
\begin{enumerate}
	\item Model - which contains data model generated by the Entity Framework (6.1) with  Code First approach,
	\item Base Engine - consist of several abstract classes or interfaces which can be used to implement user solutions, but they provide appropriate business logic,
	\item Concrete Engine - implements user logic based on previous the layer (in this paper we implemented the SURF descriptors for feature extraction and $k$-means clustering \cite{hartigan1979algorithm} for indexing),
	\item CBIR Service (for more about CBIR see \cite{Content-based-Image-Indexing-by-Data-Clustering-and-Inverse-Document-Frequency}\cite{Improved-Digital-Image-Segmentation-Based-on-Stereo-Vision-and-Mean-Shift-Algorithm}) - which is WCF service that allows to invoke engine methods as Service Oriented Applications (SOA).
	\item Endpoint (Client) - presentation layer for invoking service methods, it can be desktop, web or mobile application. All the user needs to do, is to add service reference and invoke methods.
\end{enumerate}
The agility applied in the presented approach is important, because it is not restricted to any particular implementation and it can be applied in various solutions. The architecture is presented in Fig. \ref{layer}a.

\begin{figure}
\subfigure[Class diagram for Model layer. ] {\includegraphics[width=6.2cm, height=8cm]{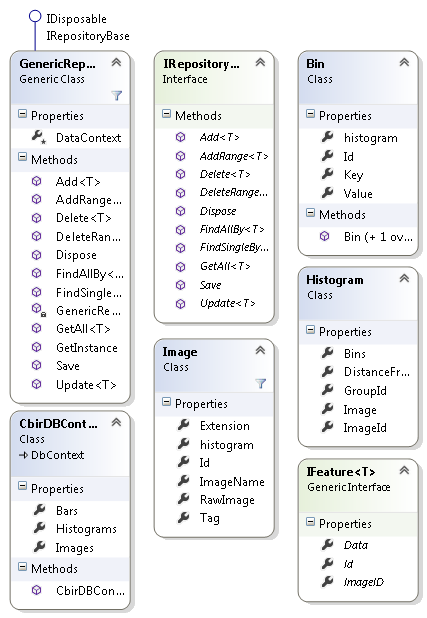}}
 \subfigure[Class diagram for CBIR Service layer.] {\includegraphics[width=6.2cm, height=8cm]{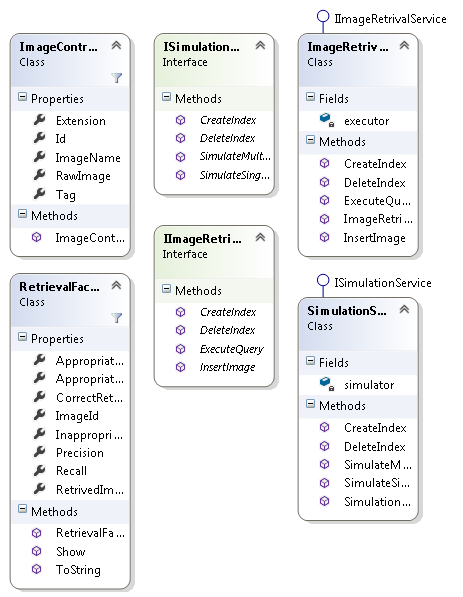}}
\caption{Class diagrams for Model and CBIR Service layers.}
\label{Model&CBIRService}
\end{figure}

The Base Engine and Concrete Engine layers are based on two design patterns, i.e. Dependency Injection (DI) and Inverse of Control (IoC) described in Section \ref{DI_IOCsub}. These patterns allowed to separate the containers and maintain the S.O.L.I.D. principles \cite{martin2006agile}. Components are composed of a set of classes or interfaces. Each component has a representation in UML (Unified Modeling Language) diagram. Figure \ref{Model&CBIRService}a shows the model layer, which contains five classes and two interfaces. Classes: $Bin$, $Histogram$ and $Image$ were generated by the Entity Framework and they correspond with the database tables presented in Fig. \ref{layer}b. Interface $IRepositoryBase$ is a generic interface which provides the basic C.R.U.D. (create, read, update and delete) operations. The methods allow to operate on any types of objects. The $GenericRepository$ implements $IRepositoryBase$ interface. In addition, the $dataContext$ field is generic, thus the concrete implementation does not contains any dependencies. The $GenericRepository$ class is based on the Singleton pattern, to create instance which user needs to use static method $GetInstance$. A very interesting interface is $IFeature$, which allows to implement any type of image features (an image descriptor).

\begin{figure}

\subfigure[Class diagram for Concrete Engine layer. ] {\includegraphics[width=5cm, height=8cm]{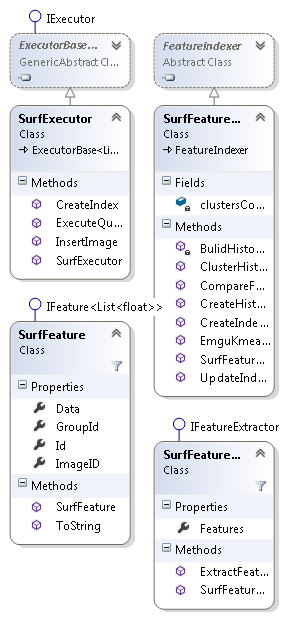}}
 \subfigure[Class diagram for Base Engine layer.] 
 {\includegraphics[width= 7cm, height=8cm] {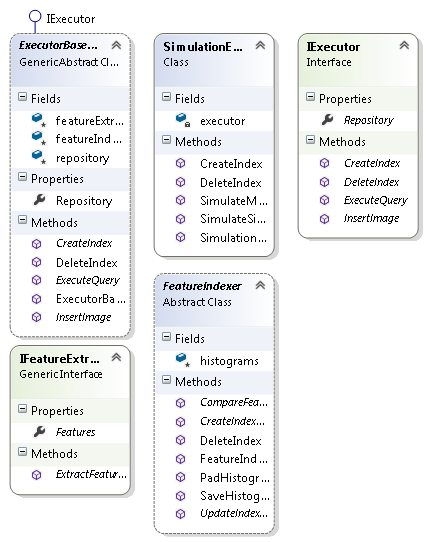}}
\caption{Class diagrams for Concrete Engine and Base Engine layers.}
\label{Concreate&EngineBase}
\end{figure}
	
	Fig. \ref{Concreate&EngineBase}b represents a class diagram for the Base Engine layer, which contains items for feature extraction ($IFeatureExtractor$), feature indexation ($Feature$ $Indexer$), operations executions ($IExecutor$, $ExecutorBase$) and simulation ($SimulationEvaluator$, $RetrievalFactors$). $IExecutor$ provides abstract methods for following operations: index creation and deletion, query execution and image insertion. It contains one property: repository. $IFeatureExtractor$ is responsible for feature extraction and contains: $Features$ property and ExtractFeatures method. The first one stores previously extracted features, the second one extracts features from image passed as a parameter. The method is only a definition, thus the class that will implement that interface must contain its own version. $FeatureIndexer$ consists of one field (histograms) which is a composition relation with the $Histogram$ class. Methods are abstract, thus they require concrete implementation in the inherited class. $ExecutorBase$ is a base class that uses the dependency injection to initiate object, that allows to inject logic. The $RetrievalFactors$ contains fields that describes the query results. The Concrete Engine layer (see Fig. \ref{Concreate&EngineBase}a) is designed and presented for simulation purposes. We implemented SURF \cite{bay2006surf}\cite{evans2009notes}\cite{grycuk2014single} as a feature extractor and $k$-means method as a indexer. Figure \ref{Model&CBIRService}b shows diagram for the CBIR Service layer. Each WCF service consists of the following items:
\begin{itemize}
	\item Interface - defines the method e.g. $IImage$ $Retrival$ $Service$,
	\item Implementation class - implements the method's body e.g. $Image$ $Retrieval$ $Service$,
	\item Contract (optional) - required to retrieve data from the service	$Image$ $Contract$.
\end{itemize}
CBIR Service component contains two services: simulation service is used to perform simulations on the created index. The most interesting methods are SimulateMultiQuery and SimulateSingleQuery. The first one returns a list of $RetrievalFactors$ and performs multi query. The second one executes a single query. The image retrieval service allows to execute queries. The difference between SimulateSingleQuery and ExecuteQuery methods are the following: ExecuteQuery returns a list of retrieved images, SimulateSingleQuery returns a list of factors ($precision$, $recall$) which allows to evaluate index efficiency. The proposed architecture was designed in .NET framework and written in C\#.
%
%
%
%
\section{Experimental Results}
Experiments were carried out using the designed architecture. The created endpoint was a desktop application with service reference to CBIR Service. Research includes experiments on various objects with background described by the SURF local interest point descriptor \cite{SIFT:Lowe:2004}. Test images were taken from the Corel database. We chose images with various types of objects. In experiments we used 90\% of each class for index creating and 10\% as query images. In Tab.\ref{tab:SimulateMultiquery} we presented the retrieved factors for each query image. Tab. \ref{fig:queryResults} shows retrieved images for a query image (the image with border). 

\begin{figure}
  \centering
  \includegraphics[height = 4cm]{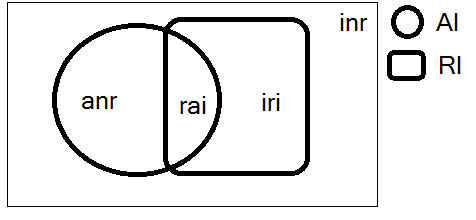}
  \caption{ Performance measures diagram. } \label{CBIRperf} 
  \end{figure}

For the purposes of the performance evaluation we use two measures; precision and recall \cite{permeskaldji2009color}. Fig. \ref{CBIRperf} shows the performance measures of the image retrieval. The $AI$ is a set of appropriate images, that should be returned as being similar to the query image. The $RI$ represents a set of returned images by the system. $Rai$ is a group of  properly returned  images. $Iri$ represents improperly returned images, $anr$ proper not returned and $inr$ improper not returned images. The presented measure allows to define $precision$ and $recall$ by the following formulas \cite{permeskaldji2009color}:
\begin{equation}
precision = \frac{|rai|}{|rai+iri|} ,
\end{equation}
\begin{equation}
recall = \frac{|rai|}{|rai+anr|} .
\end{equation}

\begin{table}%
\centering
\parbox{0.37\textwidth}{
\begin{footnotesize}
\begin{tabular}{ c | c | c | c | c | c | c | c |} 
  Image Id  &
    
     RI & 
     AI & 
     rai &
     iri &
     anr &
         \begin{sideways} Precision\end{sideways} & 
         \begin{sideways} Recall\end{sideways}  \\ 
          \hline 
      
    \hline 
    1 (1).jpg & 19 & 33 & 21 & 12 & 7 & 13 & 7 \\ 
      \hline 
    1 (2).jpg & 21 & 33 & 31 & 3 & 18 & 3 & 19 \\ 
      \hline 
    1 (20).jpg & 19 & 33 & 21 & 12 & 7 & 13 & 7 \\ 
      \hline 
    1 (21).jpg & 19 & 33 & 21 & 12 & 7 & 13 & 7 \\ 
      \hline 
    2 (1).jpg & 14 & 33 & 27 & 7 & 7 & 7 & 8 \\ 
      \hline 
    2 (10).jpg & 12 & 33 & 28 & 6 & 6 & 6 & 7 \\ 
      \hline 
    2 (11).jpg & 19 & 33 & 24 & 9 & 9 & 10 & 10 \\ 
      \hline 
    2 (17).jpg & 12 & 33 & 28 & 6 & 6 & 6 & 7 \\ 
      \hline 
    3 (1).jpg & 22 & 33 & 23 & 11 & 12 & 11 & 12 \\ 
      \hline 
    3 (10).jpg & 21 & 33 & 23 & 10 & 11 & 11 & 12 \\ 
      \hline 
    3 (11).jpg & 18 & 33 & 27 & 7 & 11 & 7 & 12 \\ 
      \hline 
    3 (15).jpg & 21 & 33 & 29 & 4 & 17 & 4 & 17 \\ 
      \hline

\end{tabular}
\end{footnotesize}
\caption{Simulation results for multi query. Measures were normalized and presented as percentage value [\%].}
\label{tab:SimulateMultiquery}
}
\qquad
\begin{minipage}[c]{0.5\textwidth}%
\centering
    \includegraphics[width=0.96\textwidth]{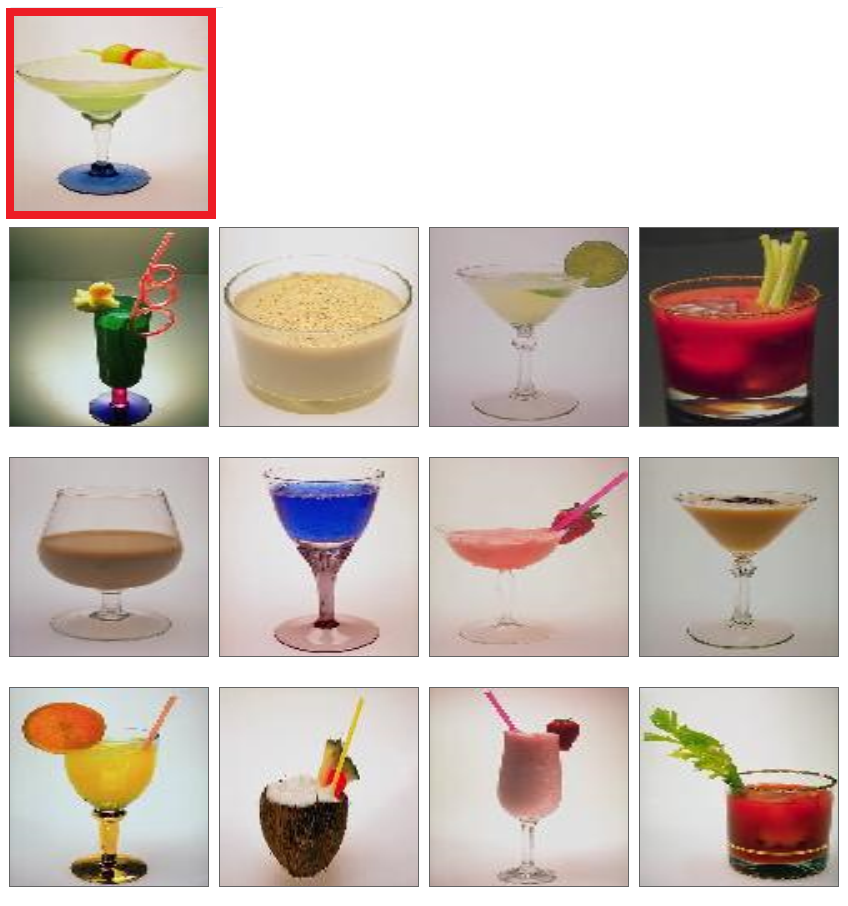}
\caption{Query results. Eighteen example images from the experiment. The image with border is the query image.}
\label{fig:queryResults}
\end{minipage}
\end{table}

\section{Final Remarks}
The presented approach is a novel architecture for storing visual data. We used SQL Server and WCF services as  a core of our method. The proposed solution for storing visual data has no dependencies with concrete implementation, thus we can simulate any CBIR method. Our approach allows creating any type of endpoint: desktop, web or mobile application. It can be used as a core back-end solution. The performed experiments proved effectiveness of our method. Our paper present a part of a larger system that allows to search and identify specific classes of objects. 

\section*{Acknowledgments}

The work presented in this paper was supported by a grant BS/MN-1-109-301/14/P "Clustering algorithms for data stream - in reference to the Content-Based Image Retrieval methods (CBIR)". The work presented in  this paper was supported partially by a grant from Switzerland through the Swiss Contribution to the enlarged European Union.
%
%
%
%
%
\bibliographystyle{splncs_srt}
\bibliography{bibref}

\end{document}